\documentclass[a4paper,fleqn,usenatbib]{mnras}

\usepackage{newtxtext,newtxmath}

\usepackage[T1]{fontenc}
\usepackage{ae,aecompl}

\usepackage{graphicx}	
\usepackage{amsmath}	
\usepackage{amssymb}	

\title[Asteroseismology of GD~133]{Precise determination of stellar parameters of the ZZ Ceti and DAZ white dwarf GD~133 through asteroseismology}

\author[J.-N. Fu et al.]{J.-N. Fu$_{,}^{1}$\thanks{E-mail: jnfu@bnu.edu.cn}
G. Vauclair$_{,}^{2,3}$
J. Su$_{,}^{4,5}$
L. Fox Machado$_{,}^{6}$
F. Colas$_{,}^{7}$
S.-L. Kim$_{,}^{8}$
\and
T. Q. Cang$_{,}^{1,2,3}$
C. Li$_{,}^{1,9}$
H. B. Niu$_{,}^{1,10}$
H. F. Xue$_{,}^{1}$
Y. Li$_{,}^{4,5}$
 X.-J. Jiang$_{,}^{9}$
\and
R. Michel$_{,}^{6}$
M. Alvarez$_{,}^{6}$
N. Dolez$_{,}^{2,3}$
L. Ma$_{,}^{10}$
A. Esamdin$_{,}^{10}$
and
J. Z. Liu$^{10}$
\\
$^1$ Department of Astronomy, Beijing Normal University, 100875 Beijing, China \\
$^2$ Universit\'e de Toulouse; UPS-OMP; IRAP; Toulouse, France \\
$^3$ CNRS; IRAP; 14 av. Edouard Belin, F-31400 Toulouse, France \\
$^4$ Yunnan Observatories, Chinese Academy of Sciences, P.O. Box 110, Kunming 650216, China \\
$^5$ Key Laboratory for the Structure and Evolution of Celestial Objects, Chinese Academy of Sciences, P.O. Box 110, Kunming 650216, China \\
$^6$ Instituto de Astronom\'ia, Observatorio Astron\'omico Nacional San Pedro M\'artir, Universidad Nacional Aut\'onoma de M\'exico,\\ 22860 Ensenada, Baja california, M\'exico \\
$^7$ IMCCE, Observatoire de Paris, UMR 8028 CNRS, F-75014 Paris, France\\
$^8$ Korea Astronomy and Space Science Institute, Daejeon 34055, Korea\\
$^9$ National Astrononomical Observatories, Chinese Academy of Sciences, 20A Datun Road, Chaoyang District, 100012 Beijing, China\\
$^{10}$ Xinjiang Astronomical Observatory, Chinese Academy of Sciences, Urumqi 830011, China
   }

\date{Accepted XXX. Received YYY; in original form ZZZ}

\pubyear{2019}

\begin{document}
\label{firstpage}
\pagerange{\pageref{firstpage}--\pageref{lastpage}}
\maketitle

\begin{abstract}
An increasing number of white dwarf stars show atmospheric chemical composition polluted by heavy elements accreted from debris disk material. The existence of such debris
disks strongly suggests the presence of one or more planet(s) whose gravitational interaction with rocky planetesimals is responsible for their disruption by tidal effect.
The ZZ Ceti pulsator and polluted DAZ white dwarf GD~133 is a good candidate for searching for such a potential planet. We started in 2011 a photometric follow-up
of its pulsations. As a result of this work in progress, we used the data gathered from 2011 to 2015 to make an 
asteroseismological analysis of GD~133, providing the star parameters from a best fit model with 
$M$/$M_{\odot}$= 0.630 $\pm$0.002 , $T_{\rm eff}$= 12400~K $\pm$70K, log($M_{\rm He}/M$)= -2.00 $\pm$0.02, 
log($M_{\rm H}/M$)= -4.50 $\pm$0.02 and determining a rotation period of $\approx$ 7 days.
\end{abstract}




\section{Introduction}

An increasing number of white dwarf stars show atmospheric chemical composition polluted by heavy elements accreted from a debris disk. The formation of the debris disk by 
the tidal disruption of rocky planetesimals suggests that one, or more, still undetected planet(s) may be orbiting the white dwarfs. Their gravitational interaction with the 
planetesimals may induce the orbital perturbations leading to the planetesimals disruption once they get close to the white dwarf tidal radius. So looking for the signature of
 such a planet
 seems justified. We have undertaken this search by starting a long-term follow-up of the ZZ Ceti pulsator and polluted white dwarf GD~133, to look for a periodic variation of the observed pulse arrival time of the oscillations.
On the other hand, carrying-out observation campaigns for the ZZ Ceti pulsators helps also to understand the 
properties of the oscillations of the target stars and precisely determine stellar parameters through asteroseismology,
which are the preconditions of the search for planets. As a preliminary result of this work, we provide in the present
paper an asteroseismological analysis of GD~133, which helps to determine the stellar parameters precisely.

\subsection{GD133 as a ZZ CETI}

GD~133 has long been considered as a non-variable DA white dwarf close to the blue edge of the ZZ~Ceti instability strip. It has been observed by various teams in an effort to 
determine precisely the location of the ZZ~Ceti instability strip blue edge. It was on our target list during an observing run in 1986 at the 1.93~m telescope of the 
Haute-Provence 
Observatory. We obtained two short runs on November 29th and December 1st. A peak was present in the Fourier-Transform at ~8138~$\mu$Hz (period $\approx$~123~s) with 
an amplitude of 4 mma. 
However, with a S/N ratio of only 2.4~$\sigma$ this peak was considered as unsignificant and this result remained unpublished. Later \citet{kepler95} 
put a 2.7 mma upper 
limit on the pulsation in GD~133. It was until GD~133 could be observed with the VLT-ULTRACAM that its pulsations were unambiguously detected by \citet{silvotti06}. 
The 
star showed three small amplitude pulsations at 120.4~s, 115.9~s and 146.9~s with amplitudes of 4.6 mma, 1.5 mma and 1.1 mma, respectively. 
The analysis of the archival Hubble Space Telescope data from the Cosmic Origins Spectrograph reveals the same 3 periods with larger amplitudes in the UV 
\citep{sandhaus16}. In addition a long period variability with a 5.2~hr period is present in these data. 
With an effective temperature of 
12600~K $\pm$ 192~K \citep{gianninas11}, GD~133 should be among the hottest ZZ Ceti pulsators defining the blue edge of the instability strip.

\subsection{GD133 as a DAZ}

GD~133 was reclassified as a DAZ after the discovery of Ca lines in its spectrum by \citet{koester05}, who derived its atmospheric parameters as 
 $T_{\rm eff}$= 12200~K, log(g)= 7.9 and an abundance [Ca/H]= -7.3. A number of subsequent analyses have redetermined the atmospheric parameters of GD~133, 
i.e. \citet{lajoie07}, \citet{koester09}, \citet{farihi10}, \citet{kawka11}, \citet{gianninas11}. The most recent 
work by \citet{xu14} revises the estimated Ca abundance as 
[Ca/H]= -7.21 $\pm$ 0.13, derives abundances for the additional heavy elements O, Mg and Si and updates the atmospheric parameters as $T_{\rm eff}$= 12600~K $\pm$ 200~K, 
in agreement with the value derived by \citet{gianninas11},
and log(g)= 8.10 $\pm$ 0.10, from which they 
derived a mass of $M/M_{\odot}$= 0.66.

The heavy elements observed in the atmosphere of GD~133 originate from matter accreted from the debris disk discovered by \citet{jura07} \citep[see also][]{jura09,farihi09}.
Their relative abundances are quite similar to the Earth bulk composition \citep{xu14}.
Once accreted on the white dwarf, they can not remain at the surface. They are rapidly mixed in the convection zone and further deeper
 in the interior by the fingering convection induced by the inverse $\mu$ gradient resulting from the accretion \citep{deal13,vauclair15,wachlin17}. Their presence in the 
atmosphere proves that the accretion of material from the debris disk is going on.  

\subsection{Is there a planet orbiting GD133?}

The existence of debris disks around white dwarfs is explained by the disruption of rocky planetesimals by tidal effect as soon as their orbit enters the tidal 
radius of the white dwarfs \citep{debes02,jura03,jura08}. These planetesimals are the remnants of the pristine planetary 
system. 
They have survived the whole central star evolution through the final 
white dwarf stage. That such planetesimals disintegration occurs is now an observational evidence after the recent discovery of the planetesimal fragments, including  
gas components, 
transiting the white dwarf WD 1145+017 \citep{vanderburg15,gansicke16,xu16,rappaport16,croll17}. 

Such planetesimals have to come close to, or cross through, the tidal radius of the white dwarf for being subsequently disintegrated. To achieve this condition, their 
presumably original circular orbits must be perturbed by gravitational interactions with massive(s) planet(s), of either Jovian or Neptunian sizes, to evolve toward 
elliptical orbits of large eccentricity so as to pass close enough to the white dwarf tidal radius. A number of scenarii of planetary system evolution through the giant 
and asymptotic giant 
branches have demonstrated that such massive planets may survive these phases of the star evolution and be able to perturb the planetesimal orbits to large eccentricity 
\citep[e.g.][]{debes02,bonsor11,debes12,mustill13,veras11,veras13,veras14a,veras14b,veras15a,veras15b,veras15,veras16a,veras16b,frewen14}.
 
Any potential planet(s) orbiting GD~133 could be detected through a periodic variation of the observed pulse arrival time of the oscillations induced by the time delay due to 
the orbital motion of the white dwarf around the barycenter of the system. 
But at the present time no planet has yet been discovered orbiting around white dwarfs by using this technique 
\citep{winget03,mullally08,mukadam13,winget15}.
 
As an example, a Jovian mass planet orbiting a $M_{*}$ = 0.6 $M_{\odot}$ white dwarf on a circular orbit with a 1 AU semi-major axis would imply a 0.8 s change in the arrival 
time $\tau$, 
on an orbital period of $\approx$ 1.2 years. For a Neptunian mass planet orbiting at 7 AU $\tau$ would be $\approx$ 0.3 s on a 24 years orbital period.
 Such orbital motions could be  
detectable through the change of the observed pulsation periods and/or in a O-C diagram. This detectability increases with both the distance of the planet to its star 
$a_{p}$ 
and the planet mass $m_{p}$ according to \citet{mullally08}:
\[
\tau = a_{p}m_{p} \sin i/M_{*}c, 
\]
where $i$ is the inclination angle of the planet orbital plane on the line of sight. 

We have selected GD~133 to study because to make such a measurement, one needs to choose a ZZ Ceti star close to the blue edge of the instability strip where they show only a few 
periods, stable in both their  periods and amplitudes.  The cool ZZ Ceti stars are unsuitable because of the growing nonlinear interactions of the pulsations with  
convection. Such interactions induce complex power spectrum with frequency combinations and amplitude variations which makes difficult a precise measurement of the periods. 
One also needs to avoid large amplitude ZZ Ceti stars because of the nonlinear effects induced by the large amplitudes.
GD~133 as a low amplitude ZZ Ceti pulsator close to the blue edge of the instability strip and as a polluted DAZ accreting heavy material from a debris disk detected through 
its IR excess is an ideal target to attempt the detection of a potential orbiting planet. In addition, the disk is seen almost edge-on, according to the disk best-fit model
described by \citet{jura09} who derived an inclination angle of $\approx$ 79~$\deg$. Assuming that the perturbing planet lies in the same plane than the disk makes 
the case most favorable for a detection. We have undertaken a long term photometric follow-up of the GD~133 pulsations which started in 2011. 
In the present paper, we use the data 
gathered up to now to determine the star parameters through asteroseismology.

The paper is organized as follows. Section 2 describes the observations obtained since 2011. Data analysis is presented in section 3. In section 4 we describe the method for 
determining the best-fit model and discuss the results. 
A summary and the conclusions are given in section 5.

\section{Observations}

We observed GD~133 during a first multisite campaign from March 2 to March 9, 2011. The campaign included three observing sites: San Pedro M\'artir (SPM) observatory 
(Mexico),
 with the 1.5~m telescope; LiJiang observatory (Yunnan, China), with the 2.4~m telescope and Pic-du-Midi (PdM) observatory (France), with the 1~m telescope. Table 1 gives the 
journal of the observations. The cycle time was 35~s at SPM and 30~s for both LiJiang and PdM. We obtained 106 hours of photometric data for a campaign total duration of 
174.6 hours, i.e. a 60.7\% duty cycle.

\begin{center}
\begin{table}
\scriptsize
\caption{Journal of observations: 2011}
\catcode `\*=\active \def*{\phantom{0}}
\baselineskip12pt
\pagestyle{empty}
\bigskip
\halign{#&\quad#\hfil&\quad#\hfil&\quad\hfil#\hfil&
\quad\hfil#\hfil&\quad#\hfil\quad\cr
\noalign{\hrule}
\noalign{\kern2pt}
\noalign{\hrule}
\noalign{\bigskip}
&Run\_Name&\quad Telescope&Date &Start\_Time &Run\_Length \cr
&          &                &(UT) &  (UTC)    &  (frames)     \cr
\noalign{\medskip}
\noalign{\hrule}
\noalign{\bigskip}
&sp0302     &      SPM 1.5-m      &   02 March & 04:37:02  &*858 \cr
&sp0304     &      SPM 1.5-m      &   04 March & 04:29:52  &*800 \cr
&lj0304     &      LiJiang 2.4-m  &   04 March & 15:33:39  &**55 \cr
&2011034    &      PdM 1-m        &   04 March & 20:29:58  &*963 \cr
&sp0305     &      SPM 1.5-m      &   05 March & 04:36:47  &*800 \cr
&lj0305     &      LiJiang 2.4-m  &   05 March & 15:38:15  &*181 \cr
&20110305   &      PdM 1-m        &   05 March & 20:24:10  &*985 \cr
&sp0306     &      SPM 1.5-m      &   06 March & 04:21:59  &*856 \cr
&lj0306     &      LiJiang 2.4-m  &   06 March & 14:08:46  &*780 \cr
&20110306   &      PdM 1-m        &   06 March & 20:33:24  &*951 \cr
&lj0307     &      LiJiang 2.4-m  &   07 March & 13:53:35  &*728 \cr
&20110307   &      PdM 1-m        &   07 March & 20:13:06  &1000 \cr
&sp0308     &      SPM 1.5-m      &   08 March & 06:22:43  &*599 \cr
&lj0308     &      LiJiang 2.4-m  &   08 March & 13:41:33  &*806 \cr
&20110308   &      PdM 1-m        &   08 March & 19:44:53  &*836 \cr
&sp0309     &      SPM 1.5-m      &   09 March & 04:16:06  &*881 \cr
\noalign{\bigskip}
\noalign{\hrule}
}

\end{table}
\end{center}

\begin{figure}
\includegraphics[width=\columnwidth]{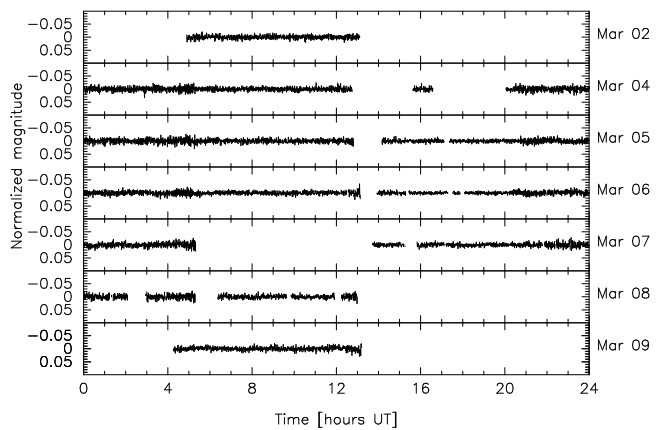}
\caption{Normalized light curve of GD~133 during the 2011 campaign. The normalized magnitude, indicated on the left side, 
 is plotted as a function of time (UT). Each panel covers a 24~h period; the date is indicated on the right side.}
\label{fig1}
\end{figure}

A second campaign was organized in 2013 from two sites. But the observations were not obtained simultaneously from the two sites. The 1.8~m telescope of the Bohyunsan 
Optical Astronomical Observatory (BOAO, Korea) made observations between March 11 and March 16, 2013. The observations were continued with the 84~cm telescope of the
San Pedro M\'artir (SPM) observatory (Mexico). Table 2 gives the journal of the 2013 observations. The observing cycle time were 35~s at BOAO and 39~s at SPM. Taking separately
 the two contributions to the campaign we obtained 26~h of photometric data during the 126.7~h of the BOAO part of the campaign, i.e. a 20.5\% duty cycle, and 37~h of 
photometric data during the 313~h of the SPM part, i.e. a 11.8\% duty cycle. For the whole campaign, including the gap between the BOAO and the SPM observations, we obtained 
a  duty cycle of 14\%.    

\begin{center}
\begin{table}
\scriptsize
\caption{Journal of observations: 2013}
\catcode `\*=\active \def*{\phantom{0}}
\baselineskip12pt
\pagestyle{empty}
\bigskip
\halign{#&\quad#\hfil&\quad#\hfil&\quad\hfil#\hfil&
\quad\hfil#\hfil&\quad#\hfil\quad\cr
\noalign{\hrule}
\noalign{\kern2pt}
\noalign{\hrule}
\noalign{\bigskip}
&Run\_Name&\quad Telescope&Date &Start\_Time &Run\_Length \cr
&          &                &(UT) &  (UTC)    &  (frames)     \cr
\noalign{\medskip}
\noalign{\hrule}
\noalign{\bigskip}
&boao0311    &        BOAO 1.8-m     &   11 March & 12:32:59  &1196 \cr
&boao0314    &        BOAO 1.8-m     &   14 March & 11:37:19  &*833 \cr
&boao0315    &        BOAO 1.8-m     &   15 March & 11:28:57  &1242 \cr
&boao0316    &        BOAO 1.8-m     &   16 March & 11:37:11  &1353 \cr
&20130323    &        SPM  84-cm     &   23 March & 04:44:38  &*664 \cr
&20130324    &        SPM  84-cm     &   24 March & 03:27:42  &1050 \cr
&20130325    &        SPM  84-cm     &   25 March & 03:02:25  &1139 \cr
&20130326    &        SPM  84-cm     &   26 March & 03:23:14  &*722 \cr
&20130328    &        SPM  84-cm     &   28 March & 07:45:40  &*332 \cr
&20130329    &        SPM  84-cm     &   29 March & 03:20:44  &*910 \cr
&20130330    &        SPM  84-cm     &   30 March & 03:40:42  &*588 \cr
&20130331    &        SPM  84-cm     &   31 March & 03:37:40  &*739 \cr
\noalign{\bigskip}
\noalign{\hrule}
}

\end{table}
\end{center}

\begin{figure}
\includegraphics[width=\columnwidth]{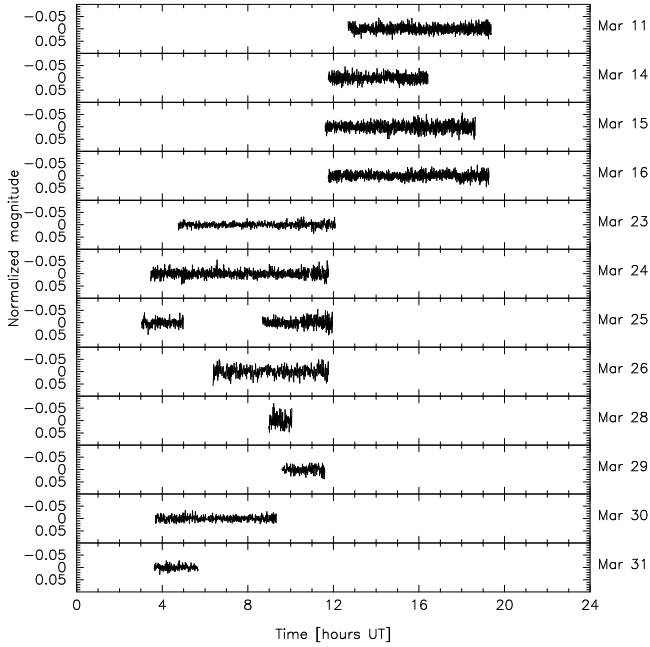}
\caption{Same as Fig.1 for the 2013 campaign.}
\label{fig2}
\end{figure}

In 2014, a third campaign was organized, including two sites in China: the Xinglong Station (XL) of the National Astronomical Observatories of China (NAOC), with the 2.16~m 
and the 85~cm telescopes, and the Nanshan observatory (Xinjiang, China), with the 1~m telescope. Data were obtained from February 27 to March 6, 2014. Table 3 gives the 
journal of the 2014 campaign. The observing cycle was 29~s on the XL 2.16~m and 28~s on the 85~cm. On the Nanshan 1~m the cycle time was adapted to the observing conditions: 
it was 44~s on March 1, 21~s on March 2, 24~s on March 3 and 20~s on March 4. We obtained 53.2~h of photometric data during the 175~h of the campaign, i.e. a duty 
cycle of 30.4\%.

\begin{center}
\begin{table}
\scriptsize
\caption{Journal of observations: 2014}
\catcode `\*=\active \def*{\phantom{0}}
\baselineskip12pt
\pagestyle{empty}
\bigskip
\halign{#&\quad#\hfil&\quad#\hfil&\quad\hfil#\hfil&
\quad\hfil#\hfil&\quad#\hfil\quad\cr
\noalign{\hrule}
\noalign{\kern2pt}
\noalign{\hrule}
\noalign{\bigskip}
&Run\_Name&\quad Telescope&Date &Start\_Time &Run\_Length \cr
&          &                &(UT) &  (UTC)    &  (frames)     \cr
\noalign{\medskip}
\noalign{\hrule}
\noalign{\bigskip}
&xl0227      &       Xinglong 2.16-m   &   27 February & 13:43:03    &*642 \cr
&xl0301      &       Xinglong 85-cm    &   01 March & 13:44:48    &*848 \cr
&ns0301      &       Nanshan 1.0-m &   01 March & 15:53:05    &*478 \cr
&xl0302      &       Xinglong 85-cm    &   02 March & 13:28:39    &*899 \cr
&ns0302      &       Nanshan 1.0-m &   02 March & 15:31:42    &*966 \cr
&ns0303      &       Nanshan 1.0-m &   03 March & 17:00:32    &*781 \cr
&xl0304      &       Xinglong 85-cm    &   04 March & 13:52:07    &*758 \cr
&ns0304      &       Nanshan 1.0-m &   04 March & 15:43:01    &*910 \cr
&xl0305      &       Xinglong 85-cm    &   05 March & 13:48:29    &*705 \cr
&xl0306      &       Xinglong 85-cm    &   06 March & 13:53:31    &*775 \cr
\noalign{\bigskip}
\noalign{\hrule}
}

\end{table}
\end{center}

\begin{figure}
\includegraphics[width=\columnwidth]{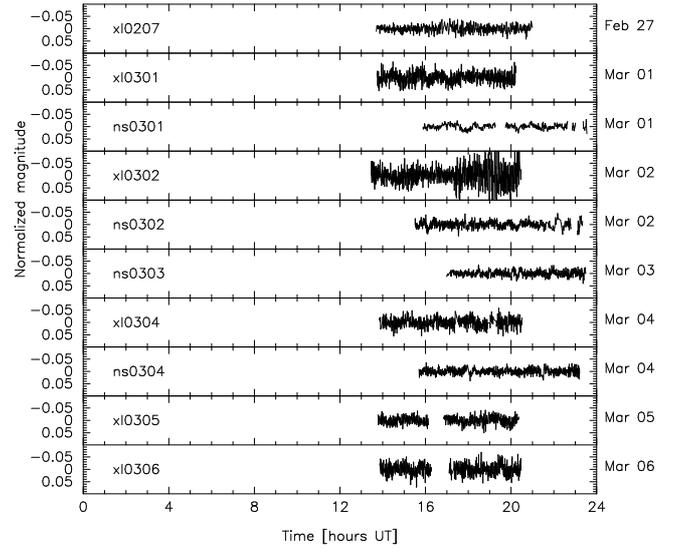}
\caption{Same as Fig.1 for the 2014 campaign.}
\label{fig3}
\end{figure}

In 2015, the multisite campaign was organized from March 13 to 23. We used the 2.16~m telescope in Xinglong and the 1~m telescope in Nanshan (China), the 1~m 
telescope in Pic-du-Midi Observatory (France) and the 84~cm 
telescope at 
San Pedro M\'artir (Mexico). Table 4 gives the journal of the 2015 observations. The observing cycles were 42~s on the Xinglong 2.16~m telescope, 22~s on the 1~m telescope at 
Pic du Midi, 42~s on the 84~cm at SPM. On the 1~m telescope of Nanshan, the observing cycle was adapted according to the observing conditions but is 30~s in average. We 
obtain a total of 72.2~h of data on the 245~h of the campaign total duration, i.e. a duty cycle of 29\%.

\begin{center}
\begin{table}
\scriptsize
\caption{Journal of observations: 2015}
\catcode `\*=\active \def*{\phantom{0}}
\baselineskip12pt
\pagestyle{empty}
\bigskip
\halign{#&\quad#\hfil&\quad#\hfil&\quad\hfil#\hfil&
\quad\hfil#\hfil&\quad#\hfil\quad\cr
\noalign{\hrule}
\noalign{\kern2pt}
\noalign{\hrule}
\noalign{\bigskip}
&Run\_Name&\quad Telescope&Date &Start\_Time &Run\_Length \cr
&          &                &(UT) &  (UTC)    &  (frames)     \cr
\noalign{\medskip}
\noalign{\hrule}
\noalign{\bigskip}
& ns0313 &  Nanshan    1-m & 13 March & 14:40:56 & 853 \cr
& ns0314 &  Nanshan    1-m & 14 March & 20:12:29 & 153 \cr
& ns0315 &  Nanshan    1-m & 15 March & 14:49:22 & 438 \cr
& ns0316 &  Nanshan    1-m & 16 March & 16:39:46 & 397 \cr
&spm317 &      SPM 84-cm & 17 March & 05:14:38 & 369 \cr
& ns0317 &  Nanshan    1-m & 17 March & 16:01:11 & 688 \cr
& xl0318 & Xinglong 2.16-m & 18 March & 12:29:04 & 633 \cr
& xl0319 & Xinglong 2.16-m & 19 March & 12:28:54 & 617 \cr
&pdm319 & PdM         1-m & 19 March & 22:49:00 & 766 \cr
& xl0320 & Xinglong 2.16-m & 20 March & 12:52:28 & 602 \cr
& xl0321 & Xinglong 2.16-m & 21 March & 12:23:40 & 479 \cr
& xl0322 & Xinglong 2.16-m & 22 March & 12:18:32 & 646 \cr
& xl0323 & Xinglong 2.16-m & 23 March & 12:14:55 & 639 \cr
\noalign{\bigskip}
\noalign{\hrule}
}

\end{table}
\end{center}


\begin{figure}
\includegraphics[width=\columnwidth]{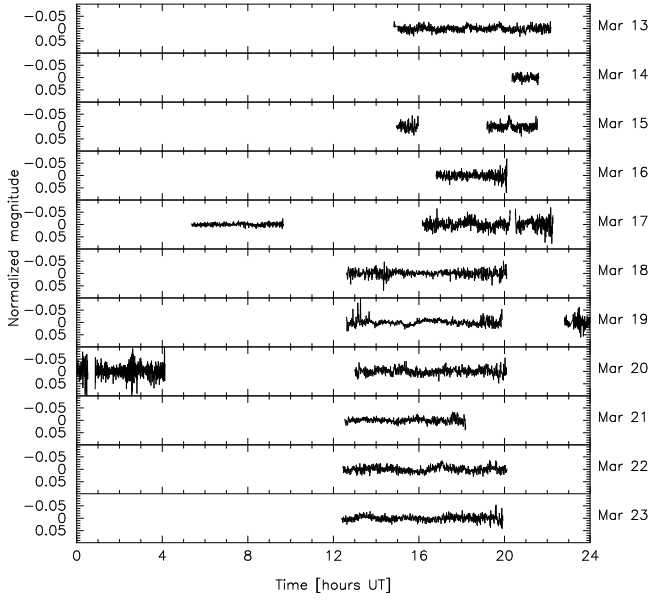}
\caption{Same as Fig.1 for the 2015 campaign.}
\label{fig4}
\end{figure}

\section{Data Analysis}

All the data were reduced with the standard IRAF routines. The IRAF AP-PHOT package was used to perform aperture photometry. Different size of apertures were optimized 
 for each night in order to obtain the minimum variance of the light curves of a check star relative to the comparison star. The light curve of GD~133 relative to the comparison star was then obtained for 
each campaign. A fifth order polynomial was applied to filter the low frequency induced by variations of the atmosphere transparency during the night. The normalized light 
curves for the 2011, 2013, 2014 and 2015 are shown in Figures 1, 2, 3 and 4, respectively.

The normalized light curve of each campaign was analyzed with the PERIOD04 software \citep{lenz05} to perform Fourier transform (FT). 
The standard prewhitening 
procedure was then applied to extract the frequencies, the amplitudes and the phases of the signals from the FT spectrum. Figures 5, 7, 8 and 9 show the Fourier transforms 
(actually the amplitude spectra) for the 2011, 2013, 2014 and 2015 campaign respectively. Figure 6 illustrates  the pre-whitening process in details on the 2011 FT.
We extract the signals from the pre-whitening process which have a signal/noise (S/N) ratio larger than 4, following the prescription given by \citet{breger93} and \citet{kuschnig97}.
 PERIOD04 estimates the internal uncertainties on the frequencies and amplitudes, which are generally underestimated. More realistic uncertainties were derived by using 
Monte-Carlo simulations as described in \citet{fu13}.
 In the case of the 2011 campaign we extracted the 8 signals listed in Table 5. The signals extracted for the 2013, 2014 and 2015 
campaigns are listed in the Tables 6, 7 and 8 respectively.  All the uncertainties quoted in the Tables are derived from the Monte-Carlo simulations.
We note that we never see any linear combinations of frequencies in our data. That is in agreement with what is expected for a low amplitude pulsator close to the blue
 edge of the ZZ Ceti instability strip. 

\begin{center}
\begin{table}
\caption{Frequencies and amplitudes of GD~133 during the 2011 campaign. The frequencies, $\nu$, are given in $\mu$Hz and the amplitudes, $A$,
in millimagnitude (mmag). The uncertainties are estimated through Monte-Carlo simulations. }
\catcode `\*=\active \def*{\phantom{0}}
\baselineskip12pt

\bigskip
\halign{# & # & # & # & # & # & # \cr
\noalign{\hrule}
\noalign{\kern2pt}
\noalign{\hrule}
\noalign{\bigskip}
&**$\nu$&**$\sigma$($\nu$)&**$A$&*$\sigma$(A)&*$P$*&*$\sigma$(P) \cr
&*($\mu$Hz)***&($\mu$Hz)*&(mmag)*&*(mmag)*&*(s)*&*(s) \cr
\noalign{\medskip}
\noalign{\hrule}
\noalign{\bigskip}
&*2821.74 & **0.34 & **0.47 &*0.15*&*354.391*&*0.042  \cr
&*5106.04 & **0.31 & **0.55 &*0.20*&*195.846*&*0.012  \cr
&*6818.48*& **0.14 & **1.33*&*0.51*&*146.660*&*0.003  \cr
&*6821.89 & **0.14 & **1.24 &*0.38*&*146.586*&*0.003  \cr
&*8308.81 & **0.16 & **1.95 &*0.43*&*120.354*&*0.002  \cr
&*8310.28 & **0.12 & **2.24 &*0.58*&*120.332*&*0.002  \cr
&*8625.25 & **0.15 & **0.72 &*0.25*&*115.938*&*0.002  \cr
&*8628.57 & **0.15 & **0.62 &*0.19*&*115.894*&*0.002  \cr
\noalign{\bigskip}
\noalign{\hrule}
}

\end{table}
\end{center}

\begin{figure}
\includegraphics[width=\columnwidth]{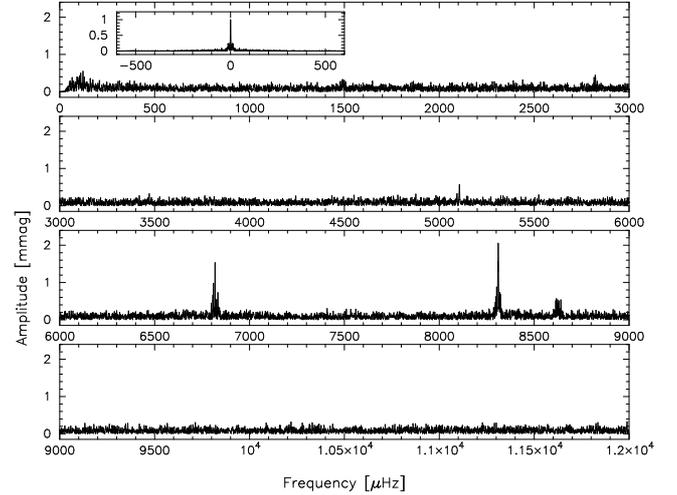}
\caption{Amplitude spectrum of the 2011 light curve. The four panels show the amplitude spectrum in units of milli-magnitude (mmag) as a function of the 
frequency in $\mu$Hz, in the frequency range 0-12000~$\mu$Hz. The corresponding window function is shown on the same scale in the insert.}
\label{fig5}
\end{figure}

\begin{center}
\begin{table}
\caption{Frequencies and amplitudes of GD~133 during the 2013 campaign.}
\catcode `\*=\active \def*{\phantom{0}}
\baselineskip12pt

\bigskip
\halign{# & # & # & # & # & # & # \cr
\noalign{\hrule}
\noalign{\kern2pt}
\noalign{\hrule}
\noalign{\bigskip}
&**$\nu$&**$\sigma$($\nu$)&**$A$&*$\sigma$(A)&*$P$*&*$\sigma$(P) \cr
&*($\mu$Hz)***&($\mu$Hz)*&(mmag)*&*(mmag)*&*(s)*&*(s) \cr
\noalign{\medskip}
\noalign{\hrule}
\noalign{\bigskip}
&*6817.40 & **0.02  & **1.10 &*0.20*&*146.6835*&*0.0004 \cr
&*6821.82 & **0.04  & **0.90 &*0.20*&*146.5884*&*0.0008 \cr
&*8308.69 & **0.02  & **1.80 &*0.10*&*120.3559*&*0.0003 \cr
&*8310.51 & **0.02  & **1.90 &*0.20*&*120.3296*&*0.0003 \cr
&*8623.56 & **0.04  & **0.80 &*0.20*&*115.9614*&*0.0005 \cr
&*8628.65 & **0.04  & **1.00 &*0.10*&*115.8930*&*0.0005 \cr
\noalign{\bigskip}
\noalign{\hrule}
}

\end{table}
\end{center}

\begin{figure}
\includegraphics[width=\columnwidth]{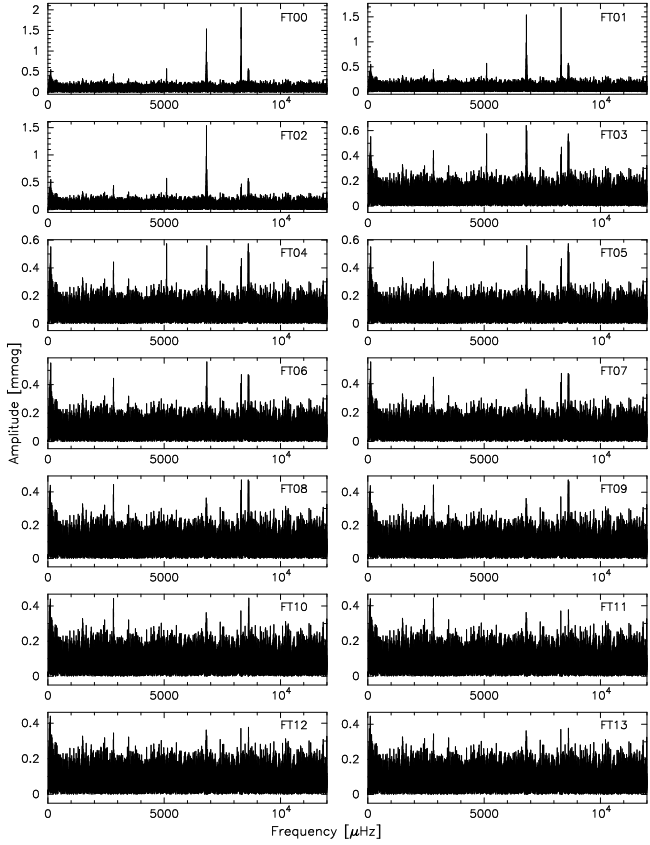}
\caption{Illustration of the prewhitening process on the 2011 amplitude spectrum.
The first panel (FT00) is the full amplitude spectrum as shown in Fig.5. Panel FT01 shows the amplitude spectrum after subtraction of the largest amplitude
 peak at 8310~$\mu$Hz. The next panel (FT02) shows the amplitude spectrum after additional subtraction of the next largest amplitude peak at 8308~$\mu$Hz, and so 
on for the subsequent panels. Note that the Y-axis scale is optimized according to the height of the largest peak in each panel.}
\label{fig6}
\end{figure}

\begin{center}
\begin{table}
\caption{Frequencies and amplitudes of GD~133 during the 2014 campaign.}
\catcode `\*=\active \def*{\phantom{0}}
\baselineskip12pt

\bigskip
\halign{# & # & # & # & # & # & # \cr
\noalign{\hrule}
\noalign{\kern2pt}
\noalign{\hrule}
\noalign{\bigskip}
&**$\nu$&**$\sigma$($\nu$)&**$A$&*$\sigma$(A)*&*$P$*&*$\sigma$(P) \cr
&*($\mu$Hz)***&($\mu$Hz)*&(mmag)*&*(mmag)*&*(s)*&*(s) \cr
\noalign{\medskip}
\noalign{\hrule}
\noalign{\bigskip}
&*6819.99 & **0.28  & **1.88 &*0.09*&*146.6278*&*0.006  \cr
&*6821.72 & **0.27  & **1.61 &*0.07*&*146.5906*&*0.006  \cr
&*8310.16 & **0.26  & **3.23 &*0.15*&*120.3346*&*0.004  \cr
\noalign{\bigskip}
\noalign{\hrule}
}

\end{table}
\end{center}

\begin{figure}
\includegraphics[width=\columnwidth]{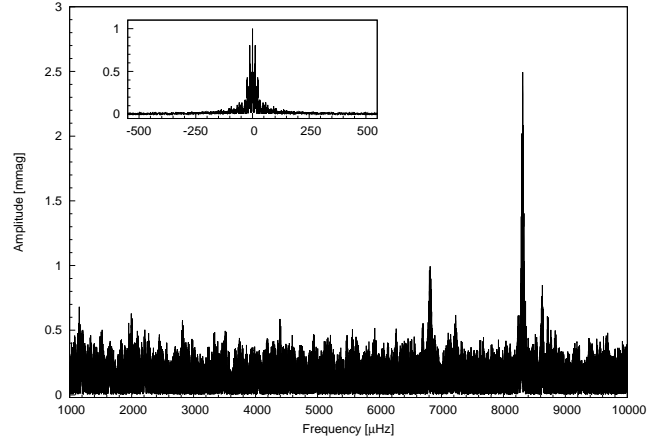}
\caption{Same as Fig.5 for the 2013 light curve. Amplitude spectrum of the 2013 light curve with the frequency range 1000-10000 $\mu$Hz.}
\label{fig7}
\end{figure}

\begin{center}
\begin{table}
\caption{Frequencies and amplitudes of GD~133 during the 2015 campaign.}
\catcode `\*=\active \def*{\phantom{0}}
\baselineskip12pt

\bigskip
\halign{# & # & # & # & # & # & # \cr
\noalign{\hrule}
\noalign{\kern2pt}
\noalign{\hrule}
\noalign{\bigskip}
&**$\nu$&**$\sigma$($\nu$)&**$A$&*$\sigma$(A)*&*$P$*&*$\sigma$(P) \cr
&*($\mu$Hz)***&($\mu$Hz)*&(mmag)*&*(mmag)*&*(s)*&*(s) \cr
\noalign{\medskip}
\noalign{\hrule}
\noalign{\bigskip}
&*6818.69 & **0.09  & **1.37 &*0.23*&*146.655*&*0.002  \cr
&*6821.51 & **0.12  & **1.49 &*0.27*&*146.595*&*0.003  \cr
&*6832.56 & **0.14  & **1.16 &*0.27*&*146.358*&*0.003  \cr
&*8308.68 & **0.06  & **1.95 &*0.21*&*120.356*&*0.001  \cr
&*8310.30 & **0.05  & **2.18 &*0.21*&*120.332*&*0.001  \cr
\noalign{\bigskip}
\noalign{\hrule}
}

\end{table}
\end{center}

\begin{figure}
\includegraphics[width=\columnwidth]{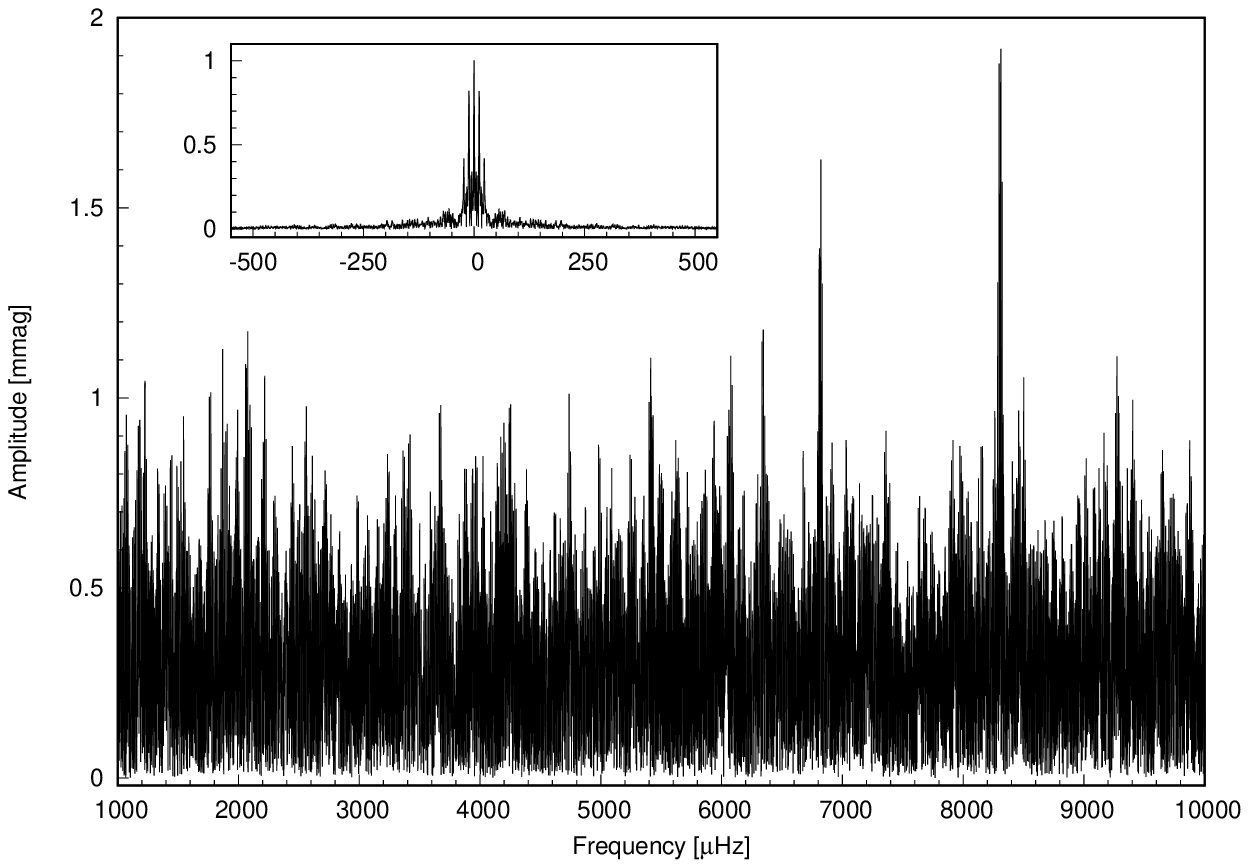}
\caption{Same as Fig. 7 for the 2014 light curve.}
\label{fig8}
\end{figure}

\begin{figure}
\resizebox{\hsize}{!}{\includegraphics{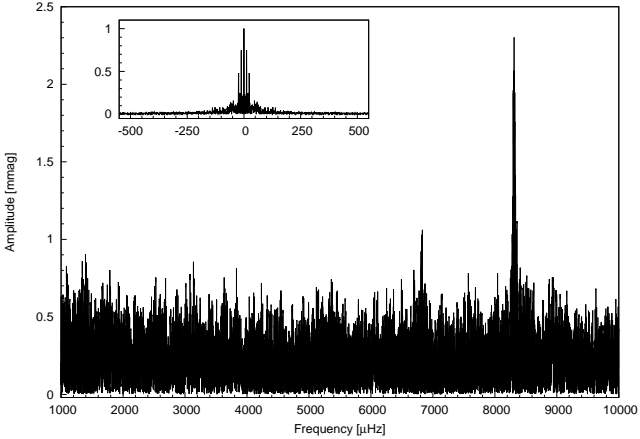}}
\vspace*{0.5cm}
\caption{Same as Fig. 7 for the 2015 light curve.}
\label{fig9}
\end{figure}

\section{Modelling}

\subsection{Theoretical models} \label{Theoretical models}

The Modules for Experiments in Stellar Astrophysics \citep[MESA, see][for details]{paxton11,paxton13,paxton15} was employed for creating carbon-oxygen white dwarf 
models. MESA creates a white dwarf model from a zero-age main sequence model, with an initial chemical composition X= 0.7, Y= 0.28 and Z= 0.02. The initial model
 evolves through main sequence and post main sequence stages with mass loss, until it remains a hot degenerate C-O core with the central temperature 
$T_{\mathrm c}$$\sim$10$^8$ K.
 The mass of the initial model is varied from 0.8 to 8 $M_{\odot}$ to get a series of C-O models with different masses. The masses of the C-O models we obtained are 
distributed from 
0.45 to 0.9 $M_{\odot}$.

In addition, an updated version of the White Dwarf Evolution Code \citep[WDEC,][]{bradley93,montgomery98} was used to construct DA white dwarf models by adjusting and relaxing the 
C-O models prepared by MESA. The masses of the helium and hydrogen layers are adjustable parameters in WDEC. It constructs H/He envelopes of given masses and chemical 
compositions
 to make DA white dwarf models. So we can conveniently obtain a model with defined parameters by choosing an initial model of given mass ($M/M_{\odot}$), setting its input 
parameters ($M_{\rm He}/M$ and $M_{\rm H}/M$) and evolving it to a specified effective temperature ($T_{\rm eff}$). Repeating the calculation, we can get a series of models with 
different stellar parameters to form a model grid for the asteroseismology analysis.

In this work, a modified version of WDEC \citep{su14} is adopted. The main modification is to incorporate the treatment of the gravitational settling into the
 code, which solves diffusion equations using the diffusion velocities and coefficients as given by \citet{thoul94}. Four elements($^1$H, $^4$He, $^{12}$C and $^{16}$O) 
are taken 
into account. All elements are assumed to be fully ionized and electrons are considered as an individual component. With the updates, the models have more physically 
realistic chemical profiles of the composition transition zones (e.g. H/He, He/C/O) than previous works \citep[for example,][]{su10}, where transition zones were approximated 
by parametric equilibrium profiles. 

A pulsation code \citep{su14}, which solves the adiabatic oscillation equations to get eigenfrequencies and eigenfunctions, is adopted to calculate the theoretical 
frequencies of the models.

 Taking into account the abundances of the elements determined by \citet{xu14}, we derive 
the metallicity of GD 133 to be approximately Z= 3.3$\times$10$^{-5}$. In this framework, we consider that the heavy elements are mixed in the whole H outer layers as a 
consequence of the fingering convection \citep{deal13}. We use an 
equivalent carbon abundance in our models to represent all of other heavy elements. 

\subsection{Asteroseismological analysis} \label{Asteroseismological analysis}

\begin{figure}
\includegraphics[width=\columnwidth]{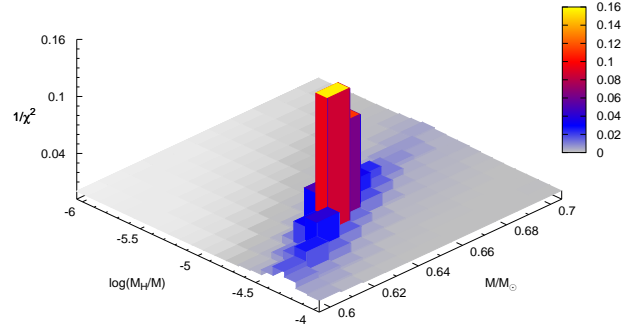}
\caption{3D $\chi^{2}$ distribution of the model grid as a function of $M/M_{\odot}$ and log($M_{H}/M$) 
For a better visibility, 1/$\chi^{2}$ is plotted with the corresponding color scale.}
\label{fig10}
\end{figure}

\begin{figure}
\includegraphics[width=\columnwidth]{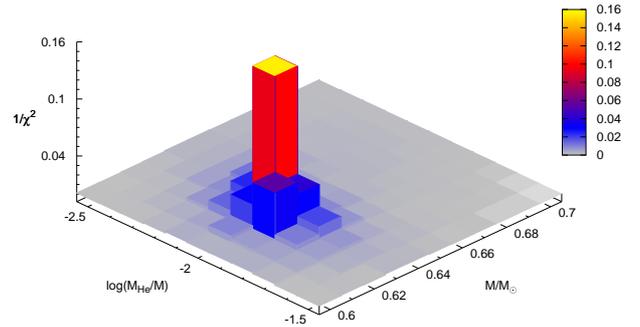}
\caption{Same as figure 10: $\chi^{2}$ as a function of $M/M_{\odot}$ and log($M_{He}/M$).}
\label{fig11}
\end{figure}

\begin{figure}
\includegraphics[width=\columnwidth]{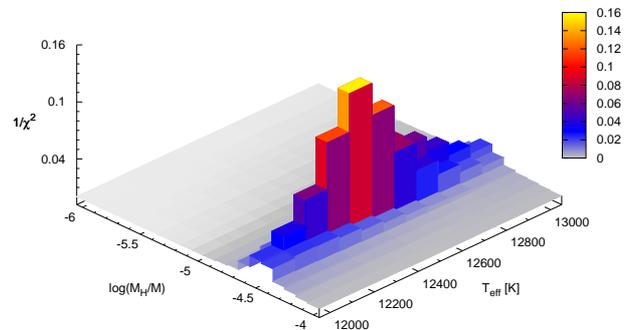}
\caption{Same as figure 10: $\chi^{2}$ as a function of $T_{\rm eff}$ and log($M_{H}/M$).}
\label{fig12}
\end{figure}

\begin{figure}
\includegraphics[width=\columnwidth]{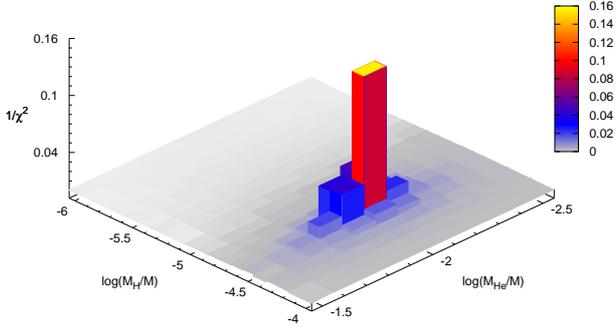}
\caption{Same as figure 10: $\chi^{2}$ as a function of  log($M_{He}/M$and log($M_{H}/M$).}
\label{fig13}
\end{figure}

The aim of the asteroseismology analysis is to find a best-fitting model by matching the theoretical periods to the observed ones. It helps us to constrain  the stellar 
parameters of GD 133. We used the frequencies derived from the 2011 multisite campaign for which we got the best duty cycle. Five periods at 115.9, 120.3, 146.7, 195.8 and 
354.4~s (the corresponding frequencies are 8625.2, 8310.3, 6818.5, 5106.0 and 2821.7~$\mu$Hz) are used to perform asteroseismological analysis. Note that we take the 
frequencies of the components whose amplitudes are the largest in the corresponding multiplets, i.e. 8625.2, 8310.3 , and 6818.5 $\mu$Hz. 
The evaluation of the
 matching degree for each model is defined as:

\begin{equation}
\chi^2\equiv\sum_{i=1}^{N}(P_{\rm ob}^{i}-P_{\rm th}^{i})^2,
\end{equation}

where $P_{\rm ob}^{i}$ are the observed periods, $P_{\rm th}^{i}$ are the theoretical periods calculated with the model and $N$ is the total number of observed periods. 
The model with 
minimum $\chi^2$ is considered to be the best-fitting model. We only calculate theoretical periods with spherical harmonic degree $\ell$= 1 and $\ell$= 2. Pulsation modes with 
higher $\ell$ values are neglected since the geometrical cancellation makes them hardly to be observed photometrically \citep[see][and references therein]{castanheira08}. We 
perform a so-called blind matching on each observed period. It lets each observed period to match the theoretical periods of either $\ell$= 1 or $\ell$= 2. Then it selects the 
theoretical period which is the closest one to the observed period and uses it as the $i$-th $P_{\rm th}$ to evaluate the $\chi^2$. One will note that the two periods 116.0 
and 120.3 s are quite close together. The period spacing between them is significantly less than the typical value for low-order g modes. This means these two periods should 
not be assigned to the same $\ell$ degree. If the period 116.0 s is $\ell$= 1 mode, the period 120.3 s must be $\ell$= 2, and vice versa. In addition, we have assumed 
implicitly that all of observed periods are $m$= 0 modes, since at this stage of analysis we lack enough evidence to determine their $m$ values. We discuss this point 
in the next section.

The spectroscopic analysis \citep{xu14} gives the atmospheric parameters of GD 133 as $T_{\rm eff}$= 12600 $\pm$ 200~K and log(g)= 8.10 $\pm$ 0.10. From log(g) and its 
uncertainty, the estimation of mass is about 0.65 $\pm$ 0.05~$M_{\odot}$. 
In order to determine the best-fit model, we computed a grid of models in the 4-D parameter space for  $ M/M_{\odot}$, $T_{\rm eff}$, $\log(M_{\rm He}/M)$ and $\log(M_{\rm H}/M)$.
The range of stellar parameters to explore is set as:
\begin{description}
\item[]0.60 $\le M/M_{\odot} \le$ 0.70,
\item[]12000 $\le T_{\rm eff} \le$ 13000~K,
\item[]-2.5 $\le \log(M_{\rm He}/M) \le$ -1.5,
\item[]-10 $\le \log(M_{\rm H}/M) \le$ -4.
\end{description}

The resolution of the grid is $\Delta_{M/M_{\odot}}$= 0.01, $\Delta_{T_{\rm eff}}$= 100~K, 
$\Delta_{\log(M_{\rm He}/M)}$= 0.1 and $\Delta_{\log(M_{\rm H}/M)}$= 0.1.

We find out the best-fitting model with a minimum $\chi^2$ of 6.40. The model parameters of the best-fitting model are 
$M/M_{\odot}$= 0.63, $T_{\rm eff}$= 12400~K, $\log(M_{\rm He}/M)$= -2, $\log(M_{\rm H}/M)$= -4.5 and the surface gravity derived from the model is log(g)= 8.09.  
The distribution of  $\chi^2$ in the 4-D parameter space can be shown as 3D-figures plotting the  $\chi^2$ values in a 2-D plane for two of the parameters with the other two parameters
fixed at the values of the best-fit model. We show four of these figures for the  $\chi^2$ distribution as function of $M/M_{\odot}$ and $log(M_{\rm H}/M)$, $M/M_{\odot}$ and $log(M_{\rm He}/M)$,
$T_{\rm eff}$ and $log(M_{\rm H}/M)$, and $log(M_{\rm He}/M)$ and $log(M_{\rm H}/M)$ in Fig.10, 11, 12 and 13 respectively.

The uncertainties on the best-fit model parameters are estimated by using the prescription derived by \citet{zhang86} which was adopted by  \citet{castanheira08} and \citet{romero12}. That is:

\[
\sigma^2 = d^2/(S-S0), 
\]

where $\sigma$ is the uncertainty of the parameter, d is chosen as the step of the parameter within the model grid. S0 is the $\chi^2$ of the best-fit model (i.e. the minimum value) and S 
is the value of $\chi^2$ for the model with the prescribed change of the parameter by the amount d, while keeping all the other parameters fixed. The best-fit model parameters and their uncertainties are:

\begin{description}
\item[]$M/M_{\odot}$ = 0.630 $\pm$ 0.002,
\item[]$ T_{\rm eff}$ = 12400~K $\pm$ 70~K,
\item[]$\log(M_{\rm He}/M)$ = -2.00 $\pm$ 0.02,
\item[]$\log(M_{\rm H}/M)$ =  -4.50 $\pm$ 0.02
\end{description}

The effective temperature and the mass of the best-fit model determined from our asteroseismic analysis are in excellent agreement with the parameters derived from the GAIA precise photometry and parallax measurement:  
$ T_{\rm eff}$ = 12259~K $\pm$ 68~K and $M/M_{\odot}$ = 0.631 $\pm$ 0.005 \citep{gentile19}. 
The hydrogen mass fraction derived for GD~133 may be compared to the hydrogen mass fractions estimated in some other DAV pulsators. In their analysis of 44 DAVs, \citet{romero12} find a range of hydrogen mass fraction 
-9.3 $\le\log(M_{\rm H}/M)\le$ -4. In the case of the well studied DAV G~117-B15A, this value is -5.9. Two other examples of hydrogen mass fraction estimated from asteroseismology of DAVs are for KUV~08368+4026 with  
 $\log(M_{\rm H}/M)$ = -4.0 \citep{li15} and HS~0507+0434B with $\log(M_{\rm H}/M)$ = -8.5 \citep{fu13}. 

The theoretical periods of the best-fitting model between 110 and 380 s are listed in Table~\ref{tab_periods}. The five observed periods are listed beside the theoretical 
periods which they are matched to. The absolute values of difference between $P_{\rm th}$ and $P_{\rm ob}$ are also listed in the table.

\subsection {Discussion on the multiplets and on the rotational splitting}

The dominant modes observed in GD~133 are the two multiplets around 8310 and 6821~$\mu$Hz. 

The first one is identified as a $\ell$= 1 mode. It is a triplet seen as a doublet during each campaign except during the 2014 campaign when only the high frequency component 
is seen. One can reasonably assume that the rotation axis of the star is perpendicular to the plane of the debris disk, which has an inclination angle of
 $\approx$ 79~$\deg$ on 
the line of sight according to \citet{jura09}. In this configuration, the star is seen almost equator-on and only the $m$= -1 and +1 components of the triplet 
are visible \citep{gizon03}. This is in excellent agreement with the fact that one sees only a doublet. The frequency splitting between the two 
components, averaged on the values measured during the 2011, 2013 and 2015 campaigns, is 1.64~$\pm$0.12$\mu$Hz, which corresponds to twice the rotational splitting $\delta\nu_{1}$= 0.82~$\mu$Hz. 
We use this value to determine the rotation period by using:
\[
\omega_{\ell,n,m}= \omega_{\ell,n} + (1-C_{\ell,n})\Omega,
\]
where $\Omega$ is the rotation frequency.   

This mode is identified in the best-fit model as the $\ell$= 1, $n$= 1 mode. To estimate the rotation period of the star one must compute the $C_{1,1}$ term since for such low 
order modes the star is not in the asymptotic regime. As a matter of fact, we find a value of 0.49996. This value is close to the value 0.5 characteristic of the asymptotic 
regime, suggesting that the mode is a trapped one. We derive a rotation period of 169~$\pm$24h ($\approx$ 7.0$\pm$1.0 days). This is a rather slow rotation compared to other white dwarfs 
with rotation period derived from asteroseismology.

\begin{table}
\centering
\caption{List of theoretical periods ($P_{\rm th}$) of the best-fitting model between 110 and 380 s. The first column is the spherical harmonic degree $\ell$ and the second 
column radial order $n$. The five observed periods ($P_{\rm ob}$) are listed beside the matched theoretical periods in the fourth column and $|P_{\rm th}-P_{\rm ob}|$ are listed 
in the fifth column.}
\begin{tabular}{ccccc}
\hline               
$\ell$ & $n$ & $P_{\rm th}$ (s) & $P_{\rm ob}$ (s) & $|P_{\rm th}-P_{\rm ob}|$ (s) \\
\hline
1& 1 & 120.88  & 120.332 & 0.55 \\
1& 2 & 196.61  & 195.846 & 0.76 \\
1& 3 & 250.43  &             &         \\
1& 4 & 315.75  &             &         \\
1& 5 & 354.85  & 354.391 & 0.46 \\
1& 6 & 378.36  &             &         \\
\hline
2& 2 & 114.85 & 115.938 & 1.09 \\
2& 3 & 144.63 & 146.660 & 2.03 \\
2& 4 & 182.66 &            &         \\
2& 5 & 205.30 &            &         \\
2& 6 & 244.61 &            &         \\
2& 7 & 265.04 &            &         \\
2& 8 & 273.75 &            &         \\
2& 9 & 310.27 &            &         \\
2&10 & 336.53 &            &         \\
2&11 & 367.38 &            &         \\
2&12 & 378.67 &            &         \\
\hline
\end{tabular}
\label{tab_periods}
\end{table}

\section{Summary and conclusions}

The white dwarf GD~133 is a promising candidate to harbor one or more planet(s) as it is surrouned by a debris disk and exhibits an atmospheric composition polluted 
 by heavy elements accreted from the debris disk. Tidal disruption of small rocky bodies, asteroid-like, is considered as the physical process resulting in the formation 
of such disks.
 Most scenarii invoke the presence of a larger mass body, of the size of a planet, in order to perturb their orbit to a large enough eccentricity and make them reaching 
the white dwarf tidal 
radius where they are disrupted. As GD~133 is also a ZZ Ceti pulsator close to the instability strip blue edge, one may search for the signature 
of such a potential planet through a periodic variation of the observed pulsation periods induced by the orbital motion.
We have undertaken in 2011 a photometric follow-up of GD~133 to search for such a potential signature. In the present paper we describe the data which have been 
gathered from 2011 to 2015. From the time series analysis we identify five independent pulsation modes, which are 
 used to make a seismic modelling of the star and derive its main parameters. The parameters of the best-fitting model are 
$M/M_{\odot}$= 0.630 $\pm$0.002, $T_{\rm eff}$= 12400~K $\pm$ 70~K, $\log(M_{\rm He}/M)$= -2.00 $\pm$ 0.02, $\log(M_{\rm H}/M)$= -4.50 $\pm$ 0.02 and the surface gravity derived from the model is log(g)= 8.09. 
The effective temperature and the mass of the best-fit model determined from our asteroseismic analysis are in excellent agreement with the parameters derived from the GAIA precise photometry and parallax measurement:  
$T_{\rm eff}$ = 12259~K $\pm$ 68~K and $M/M_{\odot}$ = 0.631 $\pm$ 0.005 \citep{gentile19}. From the rotational 
splitting we derived that GD~133 is rotating with a period of 7.0~$\pm$1.0 days.

\section*{Acknowledgements}

We thank the referee, Dr. Pierre Bergeron, for his suggestions to improve the manuscript. J.-N.F. acknowledges the support from the National Natural Science Foundation of China (NSFC) through the grants
11673003 and 11833002.
G.V. acknowledges financial support from "Programme National de Physique Stellaire" (PNPS) of CNRS/INSU, France and from Natural Science Foundation of China
(NSFC) under the grant 11673003.
J.S. acknowledges financial support from China Postdoctoral Science Foundation under the grant number 2015M570960 and support from Key Laboratory for the Structure
and Evolution of Celestial Objects, Chinese Academy of Sciences, under the grant number OP201406.
The data used in this paper were  partially acquired using the RATIR instrument, funded by the University of California (UC) and NASA Goddard Space Flight Center (GSFC),
on the  1.5 meter telescope at Observatorio Astronómico Nacional, San Pedro M\'artir, operated and maintained by OAN-SPM and IA-UNAM.
 L.F.M. and R.M. acknowledge the financial support from the UNAM under grant PAPIIT IN100918.

Based on observations made with the 2.16-m and the 85~cm telescopes
of Xinglong station operated by National
Astronomical Observatories of Chinese Academy of Sciences (China), the 1.5~m and the 84~cm of the San Pedro M\'artir observatory (Mexico),
the 2.4~m telescope of Lijiang station
operated by Yunnan Astronomical observatory of Chinese Academy of Sciences (China),
the 1~m telescope of Nanshan station operated by Xinjiang Astronomical Observatory of Chinese Academy of Sciences (China), the 1~m telescope of Pic-du-Midi observatory (France) and
the 1.8~m telescope of Bohyunsan Optical Astronomical Observatory (Korea).







\bsp	
\label{lastpage}
\end{document}